\documentclass{optica-article}
\journal{opticajournal}
\articletype{Research Article}

\usepackage{subcaption}
\usepackage{textcomp}
\usepackage{booktabs}
\usepackage[title,titletoc]{appendix}
\newcommand{\bigast}{\mathop{\vcenter{\hbox{\scalebox{1.3}{$\ast$}}}}}

\usepackage[title,titletoc]{appendix}

\usepackage{fancyhdr}

\begin{document}

\thispagestyle{fancy}
\fancyhf{} 
\lhead{Preprint submitted to Optics Express} 
\renewcommand{\headrulewidth}{0pt} 

\title{Analytical Markov Chain for Spatiotemporal Flux Evolution of the Inner Filter Effect in Fluorescent Media}

\author{Xuhui Yang\authormark{1,2} and Guofu Cao\authormark{1,*}}

\address{\authormark{1}Institute of High Energy Physics, Chinese Academy of Sciences, Beijing, 100049, China}
\address{\authormark{2}School of Physical Sciences, University of Chinese Academy of Sciences, Beijing, 100049, China}

\email{\authormark{*}caogf@ihep.ac.cn} 

\begin{abstract*}
Characterizing emission and decay time spectra in multi-component fluorescent media is essential for identifying intrinsic material properties and optimizing detectors. However, wavelength evolution from the secondary inner filter effect (IFE) distorts these observable spectra. While Monte Carlo (MC) ray-tracing can simulate this distortion, accumulating adequate tracking statistics requires long computation times, which hinders parameter optimization within high-dimensional spaces. This paper presents an analytical Markovian transport model based on spatiotemporal decoupling. A Laplace transform converts the multi-nested convolution integrals over continuous domains into a discrete Markov transition matrix, reducing the computational complexity from an exponential scale with respect to wavelength bins $N_{\lambda}$ and cascade order $n$, $\mathcal{O}(N_{\lambda}^n)$, to a linear scale, $\mathcal{O}(N_{\lambda} + n)$. The resulting algebraic solutions evaluate transient decay time spectra as a continuum superposition of Gamma wave packets and predict steady-state wavelength spectrum distortions driven by the IFE within a sub-second timescale. Validations across orthogonal and front-face spectrometer configurations show that the calculated spectra match MC simulations in lineshape. This model can serve as a fast forward engine to accelerate parameter space screening, provide early-stage detector design references, and act as a physics-constrained input for event vertex reconstruction algorithms.
\end{abstract*}


\section{Introduction}\label{sec:intro}

Fluorescent media—such as scintillators, proteins, and dyes—are central to modern spectroscopy across physics, chemistry, biology, and medicine, where precise characterization of emission spectra and decay kinetics is critical. Excitation may populate an emitting species directly or first populate a host, followed by energy transfer to a fluorophore. When absorption and emission bands overlap, secondary IFE can distort emission spectra and decay-time profiles \cite{lakowiczPrinciplesFluorescenceSpectroscopy2006,Stock:2025xkb}.

Mathematically modeling this coupled spatiotemporal transport requires solving nested convolution integrals over continuous wavelength and coordinate domains. Stochastic Monte Carlo (MC) simulations \cite{y.zhangCompleteOpticalModel2020,liRadiativeTransferLuminescent2024} can yield high precision, but sufficient statistics require many histories and substantial computation \cite{silvestriFastGPUMonte2019,maoMonteCarlobasedFullwavelength2022}. GPU implementations increase MC throughput \cite{fang_monte_2009,blyth_opticks_2025,next_collaboration_performance_2025}, while variance-reduction estimators can lower the histories needed for a specified uncertainty \cite{stolzOpticalGoniometerPaired2024}. Deterministic RTE formulations can also model fluorescence-induced wavelength shifts through wavelength-coupled source terms \cite{rasmussen_radiative_2006,sahin_radiative_2013,ma_multi-wavelength_2019}. In geometry-resolved problems, spatial, angular, wavelength, and time discretization adds computational work; the spatial grid grows with domain size, required resolution, and spatial dimension.

These computational costs constrain iterative, multi-dimensional parameter optimization during system calibration. Intrinsic absorption lengths and quantum yields vary with wavelength, whereas measured spectra already include IFE. Conventional corrections for excitation attenuation address primary IFE but do not generally correct secondary IFE in concentrated, spectrally overlapping mixtures; collection geometry can reduce this contribution \cite{foninFluorescenceDyesSolutions,kumarpanigrahiInnerFilterEffect2019,CrosstalkFluorescence2022}. Recovering intrinsic parameters therefore calls for a forward model that retains these effects. A high-dimensional fit may require thousands of forward evaluations; repeating high-statistics MC or finely discretized RTE calculations at each parameter update can therefore impose a substantial computational burden \cite{mitraOptimizationParametersCsITl2019,stolzOpticalGoniometerPaired2024}.

To overcome these challenges, we present an analytical transport model for multi-component fluorescent media without reflection or refraction. Although modern spectroscopy relies on intricate multi-fluorophore mixtures, their cascading energy-transfer chains can be decoupled into elementary radiative and non-radiative pathways. Our framework uses a representative ternary system---solvent ($\text{S}$), primary fluor ($\text{F}$), and secondary wavelength shifter ($\text{W}$)---to account for concurrent competitive absorption, non-radiative coupling, and multi-generation re-emission. The same construction also applies to other fixed component networks. Via a dual Laplace transform, the framework transforms the nested convolution integrals into a discrete Markov transition matrix, collapsing the computational complexity from exponential $\mathcal{O}(N_{\lambda}^n)$ to linear $\mathcal{O}(N_{\lambda} + n)$ for fast spatiotemporal flux evaluations.

Consequently, this analytical model serves as a high-speed alternative to repeated MC evaluations for detector calibration and future instrument design. It facilitates the separation of macroscopic IFE distortions from measured spectra and timing data, enabling phenomenological estimates of material properties such as non-radiative transfer probabilities. Validations across standard layouts—including $90^\circ$ right-angle and front-face spectroscopy—show good lineshape agreement with MC ray-tracing for common laboratory-scale cuvettes.

\section{Joint Spatiotemporal Markov Chain}\label{sec:joint_framework}

Inspired by the multi-component MC established by Y. Zhang et al. \cite{y.zhangCompleteOpticalModel2020}, this section formalizes an analytical ensemble representation of transport in multi-component media. By incorporating non-radiative cascade dynamics into a first-principles continuous model, a Laplace transform maps the coupled spatiotemporal and spectral evolution equations directly onto a discrete, low-dimensional Markovian matrix operator.

\subsection{Model Assumptions and Topological Setup}\label{subsec:assumptions}

To construct the continuous transport framework, the following physical assumptions are posited:
\par
\begingroup
\leftskip=2.5em\relax
\parindent=0pt\relax
\noindent\llap{1.\hspace{0.5em}}\textit{Homogeneity and Isotropy}: The medium is macroscopically homogeneous, and the intrinsic optical cross-sections of each individual component are spatially isotropic.\par
\noindent\llap{2.\hspace{0.5em}}\textit{Negligible Dispersion}: Optical dispersion is negligible. The radiant flux propagates at a constant group velocity $v = c/n$, enforcing a strict linear coupling between cumulative flight length and propagation time.\par
\noindent\llap{3.\hspace{0.5em}}\textit{Kasha's Rule}: For a given component, if excitation relaxes to the same emitting state, its intrinsic emission spectrum $S_{\text{em}}(\lambda)$ is independent of the initial excitation wavelength $\lambda_{\text{in}}$ \cite{AReappraisalofKashaRule2019}.\par
\noindent\llap{4.\hspace{0.5em}}\textit{Delay Independence}: For fixed material and excitation conditions, the non-radiative transfer and radiative waiting-time PDFs are independent of the incident photon wavelength. Conditional on a specified event sequence, the successive stage delays are statistically independent and follow their specified PDFs.\par
\endgroup

\begin{figure}[htbp]
\centering
\includegraphics[width=\linewidth]{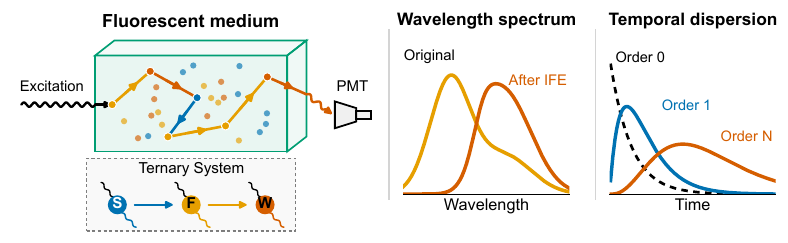}
\caption{Illustration of the ternary cascading network within the fluorescent medium.}
\label{fig:energy_transfer_topology}
\end{figure}

Based on these prerequisites, the system is modeled as a ternary cascading network consisting of a solvent ($\text{S}$), a primary fluor ($\text{F}$), and a secondary wavelength shifter ($\text{W}$). The parallel absorption competition and non-radiative serial energy transfer ($\text{S} \to \text{F} \to \text{W}$) are considered (Fig.~\ref{fig:energy_transfer_topology}).

To evaluate the collective evolution of the radiant ensemble through multiple cascade generations, we define an absorption--reemission kernel $\mathcal{I}(\lambda_{\text{in}} \to \lambda_{\text{out}}, \tau)$, representing the transition probability density that radiant energy at wavelength $\lambda_{\text{in}}$ is absorbed and, after an elapsed delay time $\tau$, survives and is re-emitted at a new wavelength $\lambda_{\text{out}}$. This kernel is constructed as a cross-section-weighted sum over all active absorption pathways:
\begin{equation}
\mathcal{I}(\lambda_{\text{in}} \to \lambda_{\text{out}}, \tau) = \sum_{i=1}^K \frac{\mu_{i}(\lambda_{\text{in}})}{\mu_{\text{tot}}(\lambda_{\text{in}})} \Phi_i(\lambda_{\text{in}} \to \lambda_{\text{out}}, \tau),
\label{eq:transfer_kernel}
\end{equation}
where $K=3$ corresponds to the constituent set $\{\text{S}, \text{F}, \text{W}\}$, $\mu_{\text{tot}}(\lambda_{\text{in}}) = \sum_i \mu_{i}(\lambda_{\text{in}})$ represents the total absorption coefficient ($\mu_i = 1/L_{i, \text{abs}}$), and $\Phi_i$ denotes the sub-kernel representing the microscopic response of component $i$ following absorption.

Conforming to the serial cascading network, these sub-kernels must account for sequential combinations of non-radiative transfers and radiative relaxations. To represent the sequential propagation of excitation through the network reaction channels, we introduce the cumulative non-radiative transfer time distribution $T_{i \to k}(\tau)$ defined for an arbitrary component pair $i \le k$:
\begin{equation}
T_{i \to k}(\tau) = \begin{cases} 
\delta(\tau), & i = k \\ 
\left( \prod_{j=i}^{k-1} P_{\text{nr}}^{j \to j+1} \right) \cdot \left[ \bigast_{j=i}^{k-1} f_{\tau}^{j \to j+1}(\tau) \right], & i < k 
\end{cases}
\label{eq:cumulative_operator_time}
\end{equation}
Within this framework, the temporal response is governed by two PDFs: $f_{\tau}^{j \to j+1}(\tau)$ for non-radiative transfer and $f_{\tau}^k(\tau)$ for the intrinsic radiative decay of component $k$. Each is a branch-conditioned waiting-time PDF: it gives the delay given that branch occurs. The corresponding whole-stage branch weights are $P_{\rm nr}^{j\to j+1}$ for transfer and $(1-P_{\rm nr}^{k\to k+1})Q^k$ for emission; multiplying each weight by its PDF gives the joint density of branch and delay. Competing-channel kinetics, including time-dependent rates, are discussed in Appendix~\ref{sec:appendix_competing_kinetics}. In the main text, $P_{\rm nr}$ and $Q$ are time-independent. Successive delays combine by continuous time-domain convolution ($\ast$), giving the unified expression for the individual sub-kernels:
\begin{equation}
\Phi_i(\lambda_{\text{in}} \to \lambda_{\text{out}}, \tau) = \sum_{k=i}^K (1-P_{\text{nr}}^{k\to k+1}) Q^k(\lambda_{\text{in}}) S_{\text{em}}^k(\lambda_{\text{out}}) \left[ T_{i \to k}(\tau) \ast f_{\tau}^k(\tau) \right],
\label{eq:phi_unified_time}
\end{equation}
where $Q^k(\lambda_{\text{in}})$ is the wavelength-dependent quantum yield assigned to component $k$, and $S_{\text{em}}^k(\lambda_{\text{out}})$ is its normalized photon emission spectrum. The factor $1-P_{\text{nr}}^{k\to k+1}$ selects termination of the non-radiative transfer chain before this yield is applied; we set $P_{\text{nr}}^{K\to K+1}=0$. 

\subsection{Excitation Kinetics and Source Configuration}\label{subsec:delay_functions}

The temporal delay profiles of the emission processes depend fundamentally on the excitation mechanism. The energy deposition mode of the radiation source determines the initial excited-state populations. While optical excitation typically populates the singlet state ($S_1$) for rapid fluorescence, ionizing particle deposition ($dE/dx$) can also generate a substantial population of long-lived triplet states ($T_1$) \cite{renReviewRecentImprovements2024}. These triplets decay slowly and can undergo triplet-triplet annihilation (TTA), which populates an emissive singlet state and contributes delayed fluorescence. To model these transient responses, we adopt a bi-exponential time delay function:
\begin{equation}
f_{\tau}(\tau) = r_{\text{fast}} \frac{1}{\tau_{\text{fast}}} e^{-\tau/\tau_{\text{fast}}} + (1 - r_{\text{fast}}) \frac{1}{\tau_{\text{slow}}} e^{-\tau/\tau_{\text{slow}}},
\label{eq:double_exponential}
\end{equation}
where $r_{\text{fast}}$ denotes the fast component fraction, while $\tau_{\text{fast}}$ and $\tau_{\text{slow}}$ represent the characteristic decay lifetimes. Evaluating this PDF via Laplace transformation yields the dipole operator $G(s)$:
\begin{equation}
G(s) = \mathcal{L}\{f_{\tau}\} = \frac{r_{\text{fast}}}{1 + s \tau_{\text{fast}}} + \frac{1 - r_{\text{fast}}}{1 + s \tau_{\text{slow}}}.
\label{eq:laplace_relaxation_operator}
\end{equation}

The weights and lifetimes are specified by the material and excitation conditions. Other normalized delay PDFs may be used through their Laplace transforms.

The configuration of the initial source state $P_0(\lambda, t, x)$ at $t=0$ reflects the spatiotemporal profile of the energy injection mode. For pure optical excitation—the standard configuration for laboratory spectroscopy instruments—the monochromatic source is instantaneous at the boundary origin as:
\begin{equation}
P_0^{\text{opt}}(\lambda, t, x) = \delta(\lambda - \lambda_{\text{in}}) \cdot \delta(t) \cdot \delta(x),
\label{eq:initial_state_optical}
\end{equation}
where $r_{\text{fast}} = 1$ is assigned to the subsequent optical radiative relaxation channels. In this mode, primary photons initiate radiative transport from $x=0$, and their initial distribution is governed purely by the optical cross-sections of the medium. \textbf{Subsequent derivations are developed primarily based on pure optical excitation.}

Conversely, for ionizing particle deposition ($dE/dx$), energy is deposited in the medium before optical transport begins. The component fractions and emission delays may be supplied directly, incorporating source-specific non-radiative transfer and relaxation:
\begin{equation}
P_0^{dE/dx}(\lambda,t,x)=\delta(x)\left[\sum_{k=1}^K Q^k S_{\text{em}}^k(\lambda)f_\tau^k(t)\right]^{dE/dx}.
\label{eq:initial_state_dedx_correct}
\end{equation}
The bracket denotes source-specific inputs: here $Q^k$ are initial photon fractions with $\sum_k Q^k=1$, and $f_\tau^k$ include the complete source-specific delay.

\subsection{Joint Spatiotemporal Transformation and Matrix Reduction}\label{subsec:recursive_equations}

\subsubsection{Coupled Transport Equation}

To model the spatiotemporal evolution of the ensemble undergoing multiple absorption-reemission cycles, we define the joint emission-event sub-probability density $P_n(\lambda, t, x)$, where $x$ parameterizes the cumulative optical path length rather than a rigid spatial coordinate. For the transition from generation $n \to n+1$, the radiant flux propagates over an intermediate distance $\Delta x = x - x'$ before undergoing a localized absorption event. Accounting for exponential uncollided attenuation and causal coupling enforced by the group velocity $v$, the local delay is constrained by $\tau = \Delta t - \Delta x/v$. The joint spatiotemporal evolution is governed by the multi-nested integral equation:
\begin{equation}
\begin{aligned}
P_{n+1}(\lambda, t, x) = & \int_{0}^{\infty} d\lambda' \int_{0}^{x} dx' \int_{0}^{t} dt' \ P_n(\lambda', t', x') \\
& \cdot \left[ \mu_{\text{tot}}(\lambda') e^{-\mu_{\text{tot}}(\lambda') (x - x')} \right] \cdot \mathcal{I}\left(\lambda' \to \lambda, t - t' - \frac{x - x'}{v}\right).
\end{aligned}
\label{eq:coupled_recursive}
\end{equation}
Unlike individual trajectory tracking in MC simulations, these integration limits reflect a continuous ensemble formulation that aggregates all valid causal pathways of the prior states. As the cascade order $n$ increases, the temporal and spatial histories become inherently intertwined through the nested convolutions.

\subsubsection{Two-Dimensional Laplace-Laplace Transformation}

To resolve the spatial and temporal convolutions, we apply a dual Laplace transform over $t$ (complex variable $s$) and $x$ (spatial variable $p$):
\begin{equation}
\tilde{P}_n(\lambda, s, p) \equiv \int_{0}^{\infty} dx \, e^{-px} \int_{0}^{\infty} dt \, e^{-st} P_n(\lambda, t, x).
\label{eq:dual_laplace_def}
\end{equation}
Applying this transform to Eq.~\eqref{eq:coupled_recursive} decouples the nested spatiotemporal integrations, mapping the continuous transport equation onto the algebraic recurrence relation:
\begin{equation}
\tilde{P}_{n+1}(\lambda, s, p) = \int_{0}^{\infty} d\lambda' \, \tilde{P}_n(\lambda', s, p) \cdot \frac{\mu_{\text{tot}}(\lambda') \tilde{\mathcal{I}}(\lambda' \to \lambda, s)}{p + \mu_{\text{tot}}(\lambda') + s/v}.
\label{eq:joint_recursive_laplace}
\end{equation}

\subsubsection{Subspace Preservation and $3 \times 3$ Matrix Reduction}

Invoking Kasha's rule, the wavelength variable $\lambda$ in Eq.~\eqref{eq:joint_recursive_laplace} appears exclusively within the intrinsic emission functions $S_{\text{em}}^{i}(\lambda)$, rendering the transport operator a mathematically separable kernel. Via mathematical induction, the nested wavelength integration is strictly preserved within a three-dimensional subspace spanned by the emission bases $\{S_{\text{em}}^{\text{S}}(\lambda), S_{\text{em}}^{\text{F}}(\lambda), S_{\text{em}}^{\text{W}}(\lambda)\}$. We introduce the subspace ansatz for the joint state vector:
\begin{equation}
\tilde{P}_n(\lambda, s, p) = \sum_{k=1}^K \tilde{C}_n^{k}(s, p) S_{\text{em}}^{k}(\lambda),
\label{eq:separable_ansatz}
\end{equation}
For optical excitation, this emission-spectrum expansion applies for $n\ge1$; the coefficients $\tilde{C}_{n}^{k}(s, p)$ are the joint spatiotemporal weights of generation $n$.

Substituting this ansatz into Eq.~\eqref{eq:joint_recursive_laplace} reduces the continuous integral mapping to a discrete $3 \times 3$ Markovian transition matrix:
\begin{equation}
\begin{pmatrix} \tilde{C}_{n+1}^{\text{S}}(s, p) \\ \tilde{C}_{n+1}^{\text{F}}(s, p) \\ \tilde{C}_{n+1}^{\text{W}}(s, p) \end{pmatrix} = 
\begin{pmatrix} 
\mathcal{M}_{\text{SS}}(s, p) & \mathcal{M}_{\text{SF}}(s, p) & \mathcal{M}_{\text{SW}}(s, p) \\ 
\mathcal{M}_{\text{FS}}(s, p) & \mathcal{M}_{\text{FF}}(s, p) & \mathcal{M}_{\text{FW}}(s, p) \\ 
\mathcal{M}_{\text{WS}}(s, p) & \mathcal{M}_{\text{WF}}(s, p) & \mathcal{M}_{\text{WW}}(s, p) 
\end{pmatrix} 
\begin{pmatrix} \tilde{C}_n^{\text{S}}(s, p) \\ \tilde{C}_n^{\text{F}}(s, p) \\ \tilde{C}_n^{\text{W}}(s, p) \end{pmatrix}.
\label{eq:matrix_equation}
\end{equation}
where the initial states are defined as $\tilde{C}_1^k(s, p) = \mathcal{A}_k(\lambda_{\text{in}}, s) / (p + \mu_{\text{tot}}(\lambda_{\text{in}}) + s/v)$ for optical excitation, and $\tilde{C}_0^{k,dE/dx}(s,p)=[Q^k\mathcal{L}\{f_\tau^k\}(s)]^{dE/dx}$ for the specified particle source.

By transforming the unified sub-kernels Eq.~\eqref{eq:phi_unified_time}, the nested time-domain convolutions reduce to algebraic products. The frequency-domain complex impedance weights $\mathcal{A}_k(\lambda, s)$ associated with each component base are expressed as:
\begin{equation}
\mathcal{A}_k(\lambda, s) = \left[ \sum_{i=1}^k \mu_i(\lambda) \tilde{T}_{i \to k}(s) \right] (1-P_{\text{nr}}^{k\to k+1}) Q^k(\lambda) G_{\text{rad}}^k(s),
\label{eq:A_components_generalized}
\end{equation}
where the non-radiative transfer operator is given by $\tilde{T}_{i \to k}(s) = \prod_{j=i}^{k-1} P_{\text{nr}}^{j \to j+1} G_{\text{nr}}^{j \to j+1}(s)$ for $i < k$, and $\tilde{T}_{i \to k}(s) = 1$ for $i = k$.

The discrete matrix elements $\mathcal{M}_{kj}(s, p)$ are defined as:
\begin{equation}
\mathcal{M}_{kj}(s, p) = \int_{0}^{\infty} S_{\text{em}}^{j}(\lambda') \frac{\mathcal{A}_k(\lambda', s)}{p + \mu_{\text{tot}}(\lambda') + s/v} \, d\lambda', \quad k,j \in \{\text{S}, \text{F}, \text{W}\}.
\label{eq:matrix_elements}
\end{equation}
This framework reduces the complexity of evaluating an $n$-order cascade over $N_{\lambda}$ discrete wavelengths from the exponential scale $\mathcal{O}(N_{\lambda}^n)$ to a linear scale $\mathcal{O}(N_{\lambda} + n)$ by leveraging low-dimensional matrix multiplications of pre-computed elements.

\subsection{Joint System Response Function}

When the cascaded radiant flux escapes the boundaries of the medium to be captured by the photocathode, it undergoes a final uncollided flight segment. In the joint Laplace-Laplace domain, this spatial escape corresponds to applying the uncollided survival factor $\frac{1}{p + \mu_{\text{tot}}(\lambda) + s/v}$. Weighting this escaping flux by the PMT quantum efficiency $QE(\lambda)$, the expected joint signal response generated at the $n$-th cascade order is derived as:
\begin{equation}
\tilde{\Psi}^{(n)}(s, p) = \int_{0}^{\infty} \sum_{k=1}^K \frac{\tilde{C}_{n}^{k}(s, p) S_{\text{em}}^{k}(\lambda)}{p + \mu_{\text{tot}}(\lambda) + s/v} \cdot QE(\lambda) \, d\lambda.
\label{eq:Psi_n_correct}
\end{equation}
Summing over all allowed generations yields the global spatiotemporal generating function of the system response: $\tilde{\Psi}_{\text{PMT}}(s, p) = \sum_{n} \tilde{\Psi}^{(n)}(s, p)$. The full transient spatiotemporal profiles can subsequently be recovered by executing a numerical two-dimensional inverse Laplace transform.

\section{Time Spectrum and Intrinsic Dispersion}\label{sec:temporal_regimes}

The joint framework can be reduced to the temporal response by integrating over cumulative optical path length. In the Laplace domain this marginalization is the one-sided limit $p\to0^+$, since $\lim_{p\to0^+}\int_0^\infty e^{-px}P_n(\lambda,t,x)\,dx=\int_0^\infty P_n(\lambda,t,x)\,dx$. The wavelength-resolved time transform is $\tilde P_n(\lambda,s,0)$, with $\mathbf C_n(s)\equiv\mathbf C_n(s,0)$ and temporal transition matrix $M_{kj}(s)\equiv\mathcal M_{kj}(s,0)$:
\begin{equation}
M_{kj}(s) = \int_{0}^{\infty} S_{\text{em}}^{j}(\lambda') \frac{\mathcal{A}_k(\lambda', s)}{\mu_{\text{tot}}(\lambda') + s/v} \, d\lambda', \quad k,j \in \{\text{S}, \text{F}, \text{W}\},
\label{eq:pure_temporal_matrix_element}
\end{equation}
which simplifies the discrete matrix recurrence relation of Eq.~\eqref{eq:matrix_equation} to $\vec{C}_{n+1}(s) = \mathbf{M}(s) \vec{C}_n(s)$, while strictly preserving the structural invariance of the emission spectrum subspace in Eq.~\eqref{eq:separable_ansatz}.

\subsection{Physical Interpretations and Gamma Wave Packets}\label{subsec:dispersion_analogy}

In the unbounded homogeneous medium, marginalizing wavelength and cumulative optical path length gives the generation-resolved emission-time sub-probability density $P_n^{\mathrm{em}}(t)=\int_0^\infty d\lambda\int_0^\infty dx\,P_n(\lambda,t,x)$.
For a unit-normalized source, its time integral is the probability weight of an emission in generation $n$; summing these weights over generations gives the expected number of emission events per source photon.
In the first-order evolution ($n=1$), one non-radiative transfer gives three successive stages: time to absorption, transfer delay, and radiative emission delay. If their waiting times are independent exponentials, their convolution has the same form as the Bateman equations for a multi-stage radioactive decay chain \cite{bateman1910solution}. At fixed incident wavelength $\lambda'$, the order-1 kernel contains the Laplace-domain factors
\begin{equation}
\left(\frac{1}{s+\gamma_{\mathrm{flight}}(\lambda')}\right)
\left(\frac{\gamma_{\mathrm{nr}}}{s+\gamma_{\mathrm{nr}}}\right)
\left(\frac{\gamma_{\mathrm{rad}}}{s+\gamma_{\mathrm{rad}}}\right),
\label{eq:single_stage_operators}
\end{equation}
where $\gamma_{\mathrm{flight}}(\lambda')=\mu_{\mathrm{tot}}(\lambda')v$ is the flight attenuation rate and $\gamma_{\mathrm{nr}}$ and $\gamma_{\mathrm{rad}}$ are the inverse mean delays of non-radiative transfer and radiative emission. A spectrally integrating PMT records the sum over wavelengths and generations, which cannot be separated by arrival time alone.

As the cascade advances to higher orders ($n=N$), the matrix multiplication $\mathbf{M}(s)^N$ generates high-order multiple poles for identical microscopic relaxation channels. For repeated solute radiative stages governed by a characteristic decay rate $\gamma$, the inverse Laplace transform of these high-order poles yields an unnormalized Gamma distribution wave packet:
\begin{equation}
\mathcal{L}^{-1} \left\{ \frac{1}{(s + \gamma)^N} \right\} = \frac{t^{N-1}}{(N-1)!} e^{-\gamma t}.
\label{eq:gamma_distribution}
\end{equation}
Under the configurations discussed in Sec.~\ref{subsec:simulation_config}, the high-order Gamma wave packet manifests in Fig.~\ref{fig:quad_numerical_validation}(a) as a systematic peak timing retardation and tail broadening.

The framework provides a physical basis for non-exponential formalisms used in complex transport systems, such as the Kohlrausch-Williams-Watts (KWW) stretched exponential, $I(t) = I_0 \exp [ - (t / \tau_{\text{str}})^\beta ]$ \cite{LeeStretchedExponential2001}, or power-law models, $I(t) = \frac{2-q}{\tau_0} [1 - (1-q)\frac{t}{\tau_0}]^{\frac{1}{1-q}}$ \cite{lakowiczFrequencyDomainFluorescenceSpectroscopy2002}. While those empirical configurations assume a global, wavelength-independent relaxation, our Markovian formalism derives an algebraic ensemble of wavelength-dependent wave packets. A multi-exponential fit ($F(t)=\sum_i A_i e^{-t/\tau_i}$) can describe independent emission channels when secondary IFE is negligible, as in suitable front-face measurements. With appreciable secondary IFE, fitted amplitudes and lifetimes generally depend on transport and need not equal intrinsic channel parameters.
\subsection{Experimental Calibration}\label{subsec:inversion_strategy}

For a specified source and collection geometry, let $\Phi(t;d)$ denote the time-resolved fluorescence response reaching the selected detector channel after transport over cumulative optical path length $d$. It includes the flight time and the additional absorption--reemission events enabled along that optical path length. The finite-length response follows from the joint system response in Eq.~\eqref{eq:Psi_n_correct}:
\begin{equation}
\Phi(t;d)=\mathcal L^{-1}_{s\to t}\mathcal L^{-1}_{p\to d}\!\left\{\sum_n\widetilde\Psi^{(n)}(s,p)\right\}.
\label{eq:finite_calibration_response}
\end{equation}
The measured signal is $S(t)=C_{\text{Noise}}(t)+[\mathcal K_{\text{inst}}\ast\Phi(\cdot;d)](t)$, where the calibrated instrument kernel $\mathcal K_{\text{inst}}$ is independent of $d$ and includes the laser-pulse time profile, PMT transit-time spread, and electronics response. Figure~\ref{fig:quad_numerical_validation}(b) shows the calculated arrival-time spectrum for a 1-cm cuvette measured in the right-angle configuration, assuming an ideal PMT timing kernel $\mathcal K_{\rm PMT}(t)=\delta(t)$ and no noise.

If reflection or refraction is included, a fixed emission point and collection window $\mathcal W$ require a weighted family of trajectories. Schematically, $\Phi_{\mathcal W}(t)=\int_{D_{\min}}^\infty\Phi_{\mathcal W}(t\mid d)\,d\Pi_{\mathcal W}(d)$, where $D_{\min}$ is the minimum travel distance, $d\Pi_{\mathcal W}$ includes collection and reflection/transmission weights, and $\Phi_{\mathcal W}(t\mid d)$ is the conditional response at that length. Wavelength, direction, and emission-history dependence is implicit in these weights and responses. This finite-boundary trajectory integral differs from the unweighted $[0,\infty)$ emission marginal; such boundary processes are outside the present model.

Calibrate $\mathcal K_{\text{inst}}$ with a reference whose fluorescence decay is known and secondary IFE is negligible, for example a dilute solution in a short-path cuvette. Use the same excitation pulse, collection geometry, and emission band as for the sample; fit the reference trace as the convolution of its known decay with $\mathcal K_{\text{inst}}$. After background subtraction, material parameters can be estimated by forward folding $\Phi$ through the measured kernel or by deconvolution. The latter is a first-kind causal convolution integral, and Tikhonov regularization can stabilize its inversion \cite{tikhonov_solutions_1977}.

Relative measurements are often used to cancel a shared instrument response. In the transform domain, dividing two signals with the same $\mathcal K_{\text{inst}}$ removes that common factor. However, when both measurements include secondary IFE, the component-dependent transport response generally remains and may not cancel. The matrix argument in the Appendix~\ref{sec:appendix_deconvolution_limits} shows that cancellation requires proportional initial component populations.

\subsection{Sub-Markovian Properties and Asymptotic Truncation Bounds}\label{subsec:Sub-Markov}

Assume $\mu_i(\lambda)\ge0$, $\mu_{\rm tot}(\lambda)=\sum_i\mu_i(\lambda)>0$, $0\le Q^i(\lambda),P_{\rm nr}^{i\to i+1}\le1$, and nonnegative, normalized emission spectra $\int S_{\rm em}^j(\lambda)d\lambda=1$. Let $\chi_i(\lambda)$ be the probability that absorption by component $i$ ultimately produces an emitted photon. Starting from $\chi_K=Q^K$, the mutually exclusive transfer and emission branches give $\chi_i=(1-P_{\rm nr}^{i\to i+1})Q^i+P_{\rm nr}^{i\to i+1}\chi_{i+1}$ for $i<K$; backward induction yields $0\le\chi_i\le1$. Since component $i$ is selected with probability $\mu_i/\mu_{\rm tot}$, summing its weighted emission probability over absorbing components gives $q(\lambda)=\sum_i(\mu_i/\mu_{\rm tot})\chi_i=\sum_k\mathcal A_k(\lambda,0)/\mu_{\rm tot}(\lambda)\le1$. For $\mathbf M_0\equiv\boldsymbol{\mathcal M}(0,0)$, integrating over the normalized emission spectrum of component $j$ gives $\sum_k(M_0)_{kj}=\int_0^\infty S_{\rm em}^j(\lambda)q(\lambda)d\lambda\le1$. Thus $\mathbf M_0$ is nonnegative and column-substochastic, so
\begin{equation}
\rho(\mathbf M_0)\le\|\mathbf M_0\|_1=\max_j\sum_k(M_0)_{kj}\le1.
\label{eq:M0_spectral_radius_bound}
\end{equation}
For $\Re s,\Re p\ge0$, each matrix element is the Laplace transform of a nonnegative event measure, so $|\boldsymbol{\mathcal M}(s,p)|\le\mathbf M_0$ entrywise; Perron-Frobenius monotonicity then gives $\rho[\boldsymbol{\mathcal M}(s,p)]\le\rho(\mathbf M_0)\le1$. For the supplied inputs, the reachable column sums are strictly below one; hence $\rho(\mathbf M_0)<1$ and the generation sum converges.

Let $N_{\max}$ denote the highest retained cascade generation; the omitted tail therefore starts at generation $N_{\max}+1$. For the zero-frequency, zero-length recurrence $\mathbf c_n=\mathbf M_0^n\mathbf c_0$, the omitted emission-event weight in the unbounded medium is
\begingroup\small
\begin{equation}
R_{N_{\max}}^{\infty}=\mathbf1^T\sum_{n=N_{\max}+1}^{\infty}\mathbf M_0^n\mathbf c_0
=\mathbf1^T\mathbf M_0^{N_{\max}+1}(\mathbf I-\mathbf M_0)^{-1}\mathbf c_0.
\label{eq:infinite_medium_tail}
\end{equation}
\endgroup
At finite optical path length $d$, let $R_{>N_{\max}}(d)$ denote the omitted detected weight from orders above $N_{\max}$. Such a detection requires the $(N_{\max}+1)$st emission before $d$. For its nonnegative path-length density $g_{N_{\max}+1}(x)$ and any $p>0$,
\begingroup\small
\begin{equation}
R_{>N_{\max}}(d)\le\int_0^d g_{N_{\max}+1}(x)\,dx
\le e^{pd}\int_0^\infty e^{-px}g_{N_{\max}+1}(x)\,dx
=e^{pd}\mathbf1^T\boldsymbol{\mathcal M}(0,p)^{N_{\max}}\mathbf C_1(0,p).
\label{eq:finite_path_tail_bound}
\end{equation}
\endgroup
since $1\le e^{p(d-x)}$ for $x\le d$. Let $\Psi_{\rm all}(d)$ be the full detected weight summed over all generations under the same source and collection conditions. With $p_d=1/d$, $N_{\max}$ is the first integer $N$ satisfying $e\,\mathbf1^T\boldsymbol{\mathcal M}(0,1/d)^N\mathbf C_1(0,1/d)/\Psi_{\rm all}(d)<\epsilon$ (Table~\ref{tab:spatial_metrics}). Positivity gives $\boldsymbol{\mathcal M}(0,1/d)\le\mathbf M_0$ and $\mathbf C_1(0,1/d)\le\mathbf C_1(0,0)$ entrywise, so the ratio is bounded by $e\,\mathbf1^T\mathbf M_0^N\mathbf C_1(0,0)/\Psi_{\rm all}(d)$. Thus the first order $N_{\max}^{(0)}(d;\epsilon)$ satisfying $e\,\mathbf1^T\mathbf M_0^N\mathbf C_1(0,0)/\Psi_{\rm all}(d)<\epsilon$ obeys $N_{\max}(d;\epsilon)\le N_{\max}^{(0)}(d;\epsilon)$; the $p_d$ criterion is tighter because it retains the finite-length weighting. The denominator here remains the finite-window detected weight; normalizing by the unbounded total would define a separate infinite-medium cutoff. Table~\ref{tab:spatial_metrics} illustrates the tighter finite-length cutoffs for the numerical inputs of Sec.~\ref{subsec:simulation_config}.

\subsection{Energy Conservation and Dissipation}\label{subsec:energy_balance}

In an unbounded medium, let $e_j=\int_0^\infty(hc/\lambda)S_{\rm em}^j(\lambda)\,d\lambda$ be the mean energy per photon emitted by component $j$, and write $\mathbf e=(e_1,\ldots,e_K)^T$. Assume that each absorption returns no more than the absorbed photon energy to the optical field on average: $\sum_k e_k\mathcal A_k(\lambda,0)/\mu_{\rm tot}(\lambda)\le hc/\lambda$. Averaging this condition over $S_{\rm em}^j$ gives $\sum_k e_k(M_0)_{kj}\le e_j$, or $\mathbf e^T\mathbf M_0\le\mathbf e^T$. Thus the mean energy of generation $n$, $E_n=\mathbf e^T\mathbf c_n$, satisfies $E_{n+1}\le E_n$.

The mean energy transferred to the medium per update of a photon from component $j$ is $h_j=e_j-\sum_k e_k(M_0)_{kj}\ge0$. With $\mathbf h^T=\mathbf e^T(\mathbf I-\mathbf M_0)$, the recurrence gives $\mathbf h^T\mathbf c_n=E_n-E_{n+1}$; summing over completed generation updates yields
\begin{equation}
\sum_{n=0}^{N}\mathbf h^T\mathbf c_n+E_{N+1}=E_0.
\label{eq:energy_balance}
\end{equation}
This is conservation of mean energy for the photons and medium. If $\rho(\mathbf M_0)<1$, then $E_{N+1}=\mathbf e^T\mathbf M_0^{N+1}\mathbf c_0\to0$, so $\sum_{n=0}^{\infty}\mathbf h^T\mathbf c_n=E_0$: the optical cascade is dissipative, with all initial photon energy ultimately transferred to the medium on average.

\section{Spatial Transport and Spectrometry Corrections}\label{sec:spatial_transport}

In macroscopic steady-state spectrometry, the photodetector integrates the incoming radiant flux over a continuous beam. Under these time-independent conditions, the microscopic temporal delays do not influence the macroscopic displacement of the radiation field. Integrating the joint spatiotemporal transport framework over the full continuous temporal domain $t \in [0, \infty)$ mathematically corresponds to taking the limit $s \to 0$ in the complex frequency domain. Enforcing this limit, the intrinsic molecular relaxation operators reduce to unity, $G_i(0) = 1$. Consequently, the joint matrix element $\mathcal{M}_{kj}(s, p)$ defined in Eq.~\eqref{eq:matrix_elements} degenerates to the pure spatial transition matrix element $\mathcal{M}_{kj}(0, p)$, where the temporal flight impedance $s/v$ vanishes from the denominator, leaving the spatial transform variable $p$ to modulate the total absorption cross-section $\mu_{\text{tot}}(\lambda')$:
\begin{equation}
\mathcal{M}_{kj}(0, p) = \int_{0}^{\infty} S_{\text{em}}^{j}(\lambda') \frac{\mathcal{A}_k(\lambda', 0)}{p + \mu_{\text{tot}}(\lambda')} \, d\lambda', \quad k,j \in \{\text{S}, \text{F}, \text{W}\}.
\label{eq:pure_spatial_matrix_element}
\end{equation}

Following this isomorphic mapping, the joint transform state vector $\tilde{P}_n(\lambda, s, p)$ and its constituent expansion weights $\tilde{C}_{n}^{k}(s, p)$ reduce to the spatial transform distributions $\tilde{\Omega}_n(\lambda, p) \equiv \tilde{P}_n(\lambda, 0, p)$ and $\vec{C}_n(p) \equiv \vec{C}_n(0, p)$, respectively. The matrix recurrence relation simplifies to $\vec{C}_{n+1}(p) = \boldsymbol{\mathcal{M}}(0, p) \vec{C}_n(p)$, preserving the structural properties of the emission spectrum subspace. At a prescribed optical path length $d$, modulating the state vector with the final uncollided escape factor yields the steady-state emission spectrum under the multi-order cascade:
\begin{equation}
\tilde{\Psi}_{\text{survive}}(\lambda, p) = \sum_{n=0}^{N_{\text{max}}} \tilde{\Omega}_n(\lambda, p) \cdot \frac{1}{p + \mu_{\text{tot}}(\lambda)}.
\label{eq:spatial_survival_spectrum}
\end{equation}

Beyond calculating the wavelength spectrum at a given optical distance, this spatial transform framework enables a direct evaluation of the statistical expectation of the cascade depth. The spatial Laplace transform of the surviving weight for the $n$-th order radiant flux is obtained by marginalizing Eq.~\eqref{eq:spatial_survival_spectrum} over the continuous wavelength domain:
\begin{equation}
\tilde{P}(n, p) = \int_{0}^{\infty} \tilde{\Omega}_n(\lambda, p) \cdot \frac{1}{p + \mu_{\text{tot}}(\lambda)} \, d\lambda.
\label{eq:P_np_marginal}
\end{equation}

Executing an inverse spatial Laplace transform isolates the surviving weight at the boundary, $P(n \vert d) = \mathcal{L}_{p \to d}^{-1}\{\tilde{P}(n, p)\}$. Consequently, for a defined optical distance $d$, the statistical expectation of the cascade order $\mathbb{E}[n \vert d]$ is derived as:
\begin{equation}
\mathbb{E}[n \vert d] = \frac{\sum_{n=0}^{N_{\max}} n \cdot P(n \vert d)}{\sum_{n=0}^{N_{\max}} P(n \vert d)},
\label{eq:cascade_expectation}
\end{equation}
which provides a quantitative metric to evaluate the average order of the secondary IFE across varying optical depths.

\subsection{Numerical Evaluation and Medium Characterization}\label{subsec:simulation_config}

To evaluate the steady-state spatial formulations and temporal profiles derived in Sec.~\ref{sec:temporal_regimes}, numerical calculations use a representative ternary liquid scintillator based on the 20-kton JUNO formulation \cite{abuslemeInitialPerformanceResults2026}. The assumed cascade comprises LAB solvent ($\text{S}$), 2.5 g/L of primary fluor PPO ($\text{F}$), and 3 mg/L of secondary wavelength shifter bis-MSB ($\text{W}$); the solvent is treated as non-luminescent, $Q^{\text{S}}=0$. The illustrative non-radiative transfer probabilities are $P_{\rm nr}^{\text{S}\to\text{F}}=0.001$ and $P_{\rm nr}^{\text{F}\to\text{W}}=0.00001$, the radiative lifetimes are $\tau_{\rm rad}^{\text{F}}=5.0$ ns and $\tau_{\rm rad}^{\text{W}}=2.0$ ns, and the illustrative transfer-delay parameters are $(\tau_{\rm fast},r_{\rm fast},\tau_{\rm slow})_{\text{S}\to\text{F}}=(1.0\,\text{ns},0.9271,30.03\,\text{ns})$ and $(\tau_{\rm fast},r_{\rm fast},\tau_{\rm slow})_{\text{F}\to\text{W}}=(10.0\,\text{ns},0.9021,200.83\,\text{ns})$. The absorption coefficients $\mu_i(\lambda)$, quantum yields $Q^i(\lambda)$, and emission spectra $S_{\rm em}^i(\lambda)$ are based on measurements summarized in \cite{y.zhangCompleteOpticalModel2020}. For illustrative ideal-window calculations, we set $QE=1$.

The wavelength-dependent absorption cross-sections exhibit a segmented profile across the spectrum. Solvent $\text{S}$ absorption dominates within $200$--$280\text{~nm}$, component $\text{F}$ governs $280$--$350\text{~nm}$, and $\text{W}$ dominates $350$--$410\text{~nm}$, beyond which competitive absorption occurs between $\text{F}$ and $\text{W}$. Below $350\text{~nm}$, the total attenuation length is compressed to $\sim 10^{-5}\text{~m}$, while within $350$--$420\text{~nm}$ it rises near-exponentially to meet the Rayleigh scattering length ($\sim 10^1\text{~m}$) near $400\text{~nm}$. Because the primary emission region of $\text{F}$ lies below $400\text{~nm}$ and heavily overlaps with these opaque absorption intervals, component $\text{F}$ experiences severe secondary IFE distortions. Conversely, above $400\text{~nm}$ the cuvette thickness ($1\text{~cm}$) is small compared with the absorption lengths of $\text{W}$, so these photons can traverse the sample with little attenuation.

\begin{figure}[htbp]
\centering
\includegraphics[width=\linewidth]{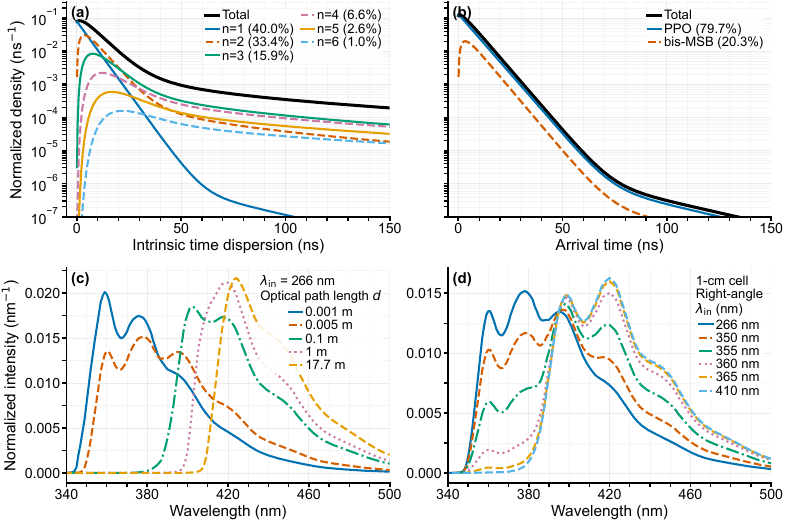}
\caption{Analytical time and wavelength spectra: (a) generation-resolved intrinsic emission-event time distributions; (b) component contributions and total arrival-time spectrum for a 1-cm cuvette in right-angle collection geometry; (c) wavelength spectra at prescribed optical path lengths $d$; (d) right-angle spectra for a 1-cm cuvette at different excitation wavelengths. Panels (a--c) use 266-nm excitation. Component contributions are shown separately and are not stacked.}
\label{fig:quad_numerical_validation}
\end{figure}

Evaluating these parameters using the precision upper bound in Sec.~\ref{subsec:Sub-Markov} with $\epsilon = 10^{-5}$ yields the conservative finite-order bounds and conditional cascade means summarized in Table~\ref{tab:spatial_metrics}. Figure~\ref{fig:quad_numerical_validation}(c) shows the 266-nm wavelength spectra at prescribed optical path lengths $d$; increasing $d$ strengthens self-absorption and shifts the spectrum from $\text{F}$-dominated peaks toward a redshifted $\text{W}$-dominated profile.

\begin{table}[htbp]
\caption{Steady-state convergence and expected cascade orders at optical path length $d$. The finite-length tail bound uses $p_d=1/d$ and $\epsilon=10^{-5}$. For comparison, $\rho(\mathbf M_0)=0.379484$ and the $\mathbf M_0$-only criterion gives $N_{\max}^{(0)}=15$ at all six lengths under the same finite-window normalization.}
\label{tab:spatial_metrics}
\centering
\small
\begin{tabular}{@{}l c c c c c c}
\toprule
$d$ (m) & $0.001$ & $0.005$ & $0.01$ & $0.1$ & $1.0$ & $17.7$ \\ \midrule
$\rho[\boldsymbol{\mathcal M}(0,1/d)]$ & $0.106519$ & $0.147282$ & $0.162632$ & $0.192458$ & $0.227927$ & $0.320577$ \\
$N_{\max}$ & $6$ & $7$ & $8$ & $9$ & $10$ & $13$ \\
$\mathbb{E}[n \vert d]$ & $1.1828$ & $1.3969$ & $1.5539$ & $2.0237$ & $2.2676$ & $2.5605$ \\ \bottomrule
\end{tabular}
\end{table}

\subsection{Analytical Modeling of Spectrometer Configurations}\label{subsec:spectrometry_modeling}

To map the steady-state spatial solutions onto laboratory cuvettes, we use the following geometric assumptions:
\par
\begingroup
\leftskip=2.5em\relax
\parindent=0pt\relax
\noindent\llap{\textbullet\hspace{0.5em}}The liquid--wall interfaces are approximated as index matched, so internal Fresnel reflection and refraction are neglected. External surface reflections are corrected using an empty reference cell and interface correction \cite{Wen:2010pba}.\par
\noindent\llap{\textbullet\hspace{0.5em}}A slight cuvette tilt reduces repeated reflections between the cell windows and the instrument optics \cite{mielenzNewReferenceSpectrophotometer1973}.\par
\noindent\llap{\textbullet\hspace{0.5em}}For the centimeter-scale cuvettes considered here, the supplied Rayleigh scattering length is much larger than $L$ (about $10\text{~m}$ near $400\text{~nm}$, compared with a $1\text{~cm}$ cuvette). The probability of scattering during one cell traversal is therefore small, so we neglect it and use ballistic paths.\par
\noindent\llap{\textbullet\hspace{0.5em}}The detector is sufficiently far from the cuvette that its collection solid angle $\Delta\Omega$ is approximately uniform over the viewed region.\par
\endgroup

These assumptions define the idealized spectrometer configurations used below.

\subsubsection{UV-Vis Spectrophotometer Configuration}\label{subsubsec:uv_vis_model}

In standard dual-beam ultraviolet-visible (UV-Vis) spectrophotometry across a thickness $d$, additional flux from the secondary IFE distorts the absorbance profile.

In this dual-beam configuration, the reference path yields a strong signal, $S_{\text{ref}} \approx I_0(\lambda_{\text{in}}) QE(\lambda_{\text{in}})$, whose intensity far exceeds the background noise. The macroscopic sample signal $S_{\text{sample}}$ gathers the unattenuated ballistic beam and the forward-emitted secondary fluorescence collected over the detection solid angle $\Delta \Omega / 4\pi$. The explicit net absorbance model is derived as:
\begin{equation}
A_{\text{mix}}(\lambda_{\text{in}}, d) = -\log_{10} \left[ e^{-\mu_{\text{tot}}(\lambda_{\text{in}}) d} + \frac{\Delta \Omega}{4\pi} \int_{0}^{\infty} \frac{\Psi_{\text{fluo}}(\lambda, \lambda_{\text{in}}, d) QE(\lambda)}{QE(\lambda_{\text{in}})} \, d\lambda + \eta_{\text{N}} \right],
\label{eq:net_absorbance}
\end{equation}
where $\Psi_{\text{fluo}}(\lambda, \lambda_{\text{in}}, d) = \sum_{n \geq 1}\int_0^d \Omega_n(\lambda,x)e^{-\mu_{\text{tot}}(\lambda)(d-x)}\,dx$ defines the secondary fluorescence that survives the final flight to the prescribed boundary. It is the inverse transform of Eq.~\eqref{eq:spatial_survival_spectrum} with the primary beam excluded. The normalized background noise is defined as $\eta_{\text{N}} = S_{\text{N}} / [I_0(\lambda_{\text{in}}) QE(\lambda_{\text{in}}) T_{\text{reflec}}]$, where $S_{\text{N}}$ is the background noise, $I_0(\lambda_{\text{in}})$ is the incident flux, and $T_{\text{reflec}}$ represents the boundary transmittance. Due to the secondary IFE term within the logarithm, the apparent absorbance systematically shifts beneath the true physical absorption baseline ($A_{\text{mix}} < A_{\text{true}}$), leading to an overestimation of the medium transmission length if inverted via the uncorrected Beer-Lambert law.

\subsubsection{Right-Angle (RA) Fluorometry Configuration}\label{subsubsec:orthogonal_model}

A right-angle (RA) fluorometer excites the sample along the $X$ axis and collects fluorescence along the perpendicular $Y$ axis. The excitation slit selects source depths $x\in[x_1,x_2]$. Let $\mu_0=\mu_{\rm tot}(\lambda_{\rm in})$ and $\mathbf A_{\rm em}(\lambda_{\rm in},0)=(\mathcal A_{\rm F},\mathcal A_{\rm W})^T$ be the first-generation source in the emitting subspace $\{\text{F},\text{W}\}$. Integrating the attenuated incident excitation along the slit-selected source-depth interval gives
\begin{equation}
\mathbf C_{\rm start}(\lambda_{\rm in})=I_0(\lambda_{\rm in})\int_{x_1}^{x_2}e^{-\mu_0x}\mathbf A_{\rm em}(\lambda_{\rm in},0)\,dx
=I_0(\lambda_{\rm in})\frac{e^{-\mu_0x_1}-e^{-\mu_0x_2}}{\mu_0}\mathbf A_{\rm em}(\lambda_{\rm in},0).
\label{eq:orthogonal_source_term}
\end{equation}
For a centered excitation ray, the transverse distance to either side face is $L/2$. In the following equation, $\mathcal L^{-1}_{p\to L/2}$ denotes the inverse path-length transform evaluated at that distance. Let $\mathbf S_{\rm em}(\lambda)=(S_{\rm em}^{\rm F}(\lambda),S_{\rm em}^{\rm W}(\lambda))^T$, and let $\mathbf M_{\rm em}$ be the $\{\text{F},\text{W}\}$ block of the transition matrix. For compactness, write $\mu_{\rm out}=\mu_{\rm tot}(\lambda_{\rm out})$. The measured right-angle spectrum is obtained by applying the final-flight survival factor, summing all re-emission generations, and inverting the path-length transform:
\begingroup\small
\begin{equation}
S_{\rm RA}(\lambda_{\rm out},\lambda_{\rm in})=I_0(\lambda_{\rm in})\eta_{\rm geo}(\lambda_{\rm out})\left[\frac{e^{-\mu_0x}}{\mu_0}\right]_{x_2}^{x_1}\mathbf S_{\rm em}^{T}(\lambda_{\rm out})\mathcal L^{-1}_{p\to L/2}\!\left\{\frac{\left[\mathbf I-\mathbf M_{\rm em}(0,p)\right]^{-1}\mathbf A_{\rm em}(\lambda_{\rm in},0)}{p+\mu_{\rm out}}\right\}.
\label{eq:measured_orthogonal_spectrum}
\end{equation}
\endgroup
For paraxial collection, $\eta_{\rm geo}(\lambda)=\Delta\Omega_{\rm det}QE(\lambda)/(4\pi n^2)$ accounts for the detector solid angle and response. The illustrative spectra set $\eta_{\rm geo}=1$, so no additional wavelength-dependent collection or detector factor is applied. Here $\sum_{n=0}^{\infty}\mathbf M_{\rm em}^n=(\mathbf I-\mathbf M_{\rm em})^{-1}$, and Eq.~\eqref{eq:orthogonal_source_term} supplies the excitation-depth integral once. Figure~\ref{fig:quad_numerical_validation}(d) shows the 1-cm RA spectra for different excitation wavelengths. Around $350\,\text{nm}$, the F and W absorption coefficients cross; changing $\lambda_{\rm in}$ across this region changes the initial F/W source balance and hence the measured spectrum.

\subsubsection{Front-Face (FF) Spectroscopy Configuration}\label{subsubsec:front_face_model}

In front-face (FF) spectroscopy, excitation and collection use the same cuvette face. We first take normal incidence and normal return. Practical instruments may tilt the detector by $30^\circ$--$45^\circ$ to reject specular reflection; at such an angle, the distance from each source depth to the viewed face changes, and the aperture views a depth-dependent part of the illuminated volume. That setup would require its angle and aperture to be included in the spatial integral.

For the normal-return model, excitation reaching depth $x_0$ is attenuated by $e^{-\mu_0x_0}$, and fluorescence returns over distance $x_0$. The incident and return segments therefore have a combined length $2x_0$. Using the same emission-space transition matrix as in the RA calculation gives the finite-cell FF spectrum directly:
\begingroup\small
\begin{equation}
S_{\rm FF}(\lambda_{\rm out},\lambda_{\rm in})=I_0(\lambda_{\rm in})\eta_{\rm geo}(\lambda_{\rm out})\!\mathbf S_{\rm em}^{T}(\lambda_{\rm out})\!\int\nolimits_0^L\!e^{-\mu_0x_0}\mathcal L^{-1}_{p\to x_0}\!\bigl\{\frac{\left[\mathbf I-\mathbf M_{\rm em}(0,p)\right]^{-1}\mathbf A_{\rm em}(\lambda_{\rm in},0)}{p+\mu_{\rm out}}\bigr\}\,dx_0.
\label{eq:finite_front_face_source}
\end{equation}
\endgroup
The finite-cell expression in Eq.~\eqref{eq:finite_front_face_source} approaches its thick-cell limit when $\mu_0L\gg1$, or equivalently when $L$ is much larger than the incident absorption length $1/\mu_0$. The incident-excitation attenuation then makes the contribution from depths beyond $L$ negligible, so the upper integration limit can be extended to infinity; the resulting integral is the spatial Laplace transform of the return response evaluated at $p=\mu_0$.

The unidirectional FF model assumes that successive emitted photons return toward the entrance. If the emitted direction is instead sampled bidirectionally, the source and intermediate depths become coupled. For a representative $\mathtt{R}\to\mathtt{L}\to\mathtt{R}$ sequence, the total optical path length is $2x_0-2x_1+2x_2$, and the spatial survival factors give the geometry integral
\begin{equation}
\int_{0}^{L} dx_0 \int_{0}^{x_0} dx_1 \int_{x_1}^{L} dx_2 e^{-\mu_0 x_0} e^{-\mu_1 (x_0-x_1)} e^{-\mu_2 (x_2-x_1)}e^{-\mu_3 x_2}.
\label{eq:dyson_series}
\end{equation}
Equation~\eqref{eq:dyson_series} shows only the geometric factors of one bidirectional FF history; component and spectral weights are omitted. The prescribed unidirectional approximation is checked against isotropic MC simulations.

\subsubsection{Cross-Validation across Transport Models}\label{subsubsec:mc_validation}

\begin{figure}[p]
\centering
\includegraphics[width=\linewidth,height=0.90\textheight,keepaspectratio]{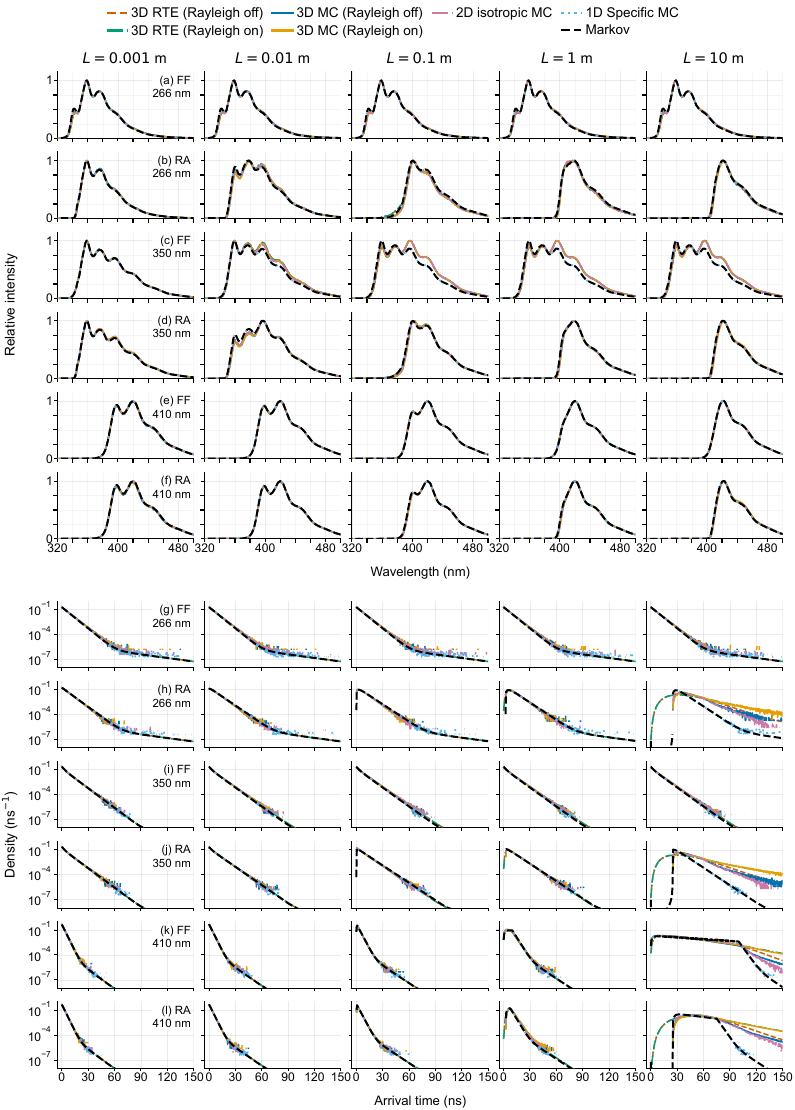}
\caption{Transport-model comparisons across five cuvette widths $L$ (columns). Spectral row pairs (a,b), (c,d), and (e,f) use 266-, 350-, and 410-nm excitation; arrival-time pairs (g,h), (i,j), and (k,l) follow the same order, with FF above RA in each pair. The legend identifies 1D Specific MC and its prescribed-path Markov solution, 2D isotropic MC, and 3D MC/RTE with Rayleigh scattering omitted or included.}
\label{fig:all_size_comparison}
\end{figure}

Figure~\ref{fig:all_size_comparison} compares the Markov model with six transport calculations. The common cuvette width is $L=0.001,0.01,0.1,1,$ or $10$~m; the 2D domain is $L\times L$, and the 3D domain is $L\times L\times10L$. A centered pencil beam enters normal to the front face. Because FF measurements seek to limit secondary IFE, collection is restricted to a centered aperture of width $w_{\rm FF}=\min[\max(10/\mu_{\rm tot}(\lambda_{\rm in}),1\,\mathrm{mm}),L]$: direction changes during a random walk can return deeply penetrating photons outside the entrance aperture, so this ten-absorption-length scale rejects much of that contribution. The 1-mm lower bound represents a practical minimum aperture, while $L$ is the fully open cuvette face. The RA window is $L\times L$. The 2D windows are the corresponding line segments. A photon is collected only when it exits through the selected window; escape through any other face terminates its contribution. Reflection and refraction are omitted.

The calculations use the following propagation rules:
\par
\begingroup
\leftskip=2.5em\relax
\parindent=0pt\relax
\noindent\llap{\textbullet\hspace{0.5em}}\textit{1D Specific MC}: Histories follow the prescribed rectilinear trajectories. In FF, each emitted photon returns along the incident axis. In RA, excitation light travels to its first absorption; the resulting fluorescence turns through $90^\circ$, and later absorption--reemission events preserve that rectilinear direction.\par
\noindent\llap{\textbullet\hspace{0.5em}}\textit{2D/3D isotropic MC}: After each fluorescence emission, a new direction is sampled uniformly in the plane (2D) or on the sphere (3D). A history ends at its next absorption or when it exits a cuvette face; only photons exiting through the selected collection window are counted.\par
\noindent\llap{\textbullet\hspace{0.5em}}\textit{3D RTE}: We solve the wavelength-coupled RTE using the formulation of Ma et al.~\cite{ma_multi-wavelength_2019} and a Cartesian upwind finite-volume mesh. Angular integration uses four polar and eight azimuthal directions. Eight kinetic state variables track the fast and slow terms for the $\mathrm{S}\to\mathrm{F}$ pathways ending in $\mathrm{F}$ or $\mathrm{W}$ emission, the fast and slow $\mathrm{F}\to\mathrm{W}$ non-radiative transfer, and radiative decay of $\mathrm{F}$ and $\mathrm{W}$. A second-order backward differentiation formula (BDF2) advances the coupled transport and kinetics, with a backward-Euler first step.\par
\noindent\llap{\textbullet\hspace{0.5em}}\textit{Markov}: Equations~\eqref{eq:measured_orthogonal_spectrum} and \eqref{eq:finite_front_face_source}, together with the joint response in Sec.~\ref{sec:joint_framework}, evaluate the same prescribed rectilinear trajectories as 1D Specific MC.\par
\endgroup

The 1D and 2D MC runs condition on successful first fluorescence; the 3D MC runs sample incident-photon histories.

When Rayleigh scattering is included, the Rayleigh scattering coefficient is the reciprocal scattering length, $\mu_{\rm R}(\lambda)=1/\ell_{\rm R}(\lambda)$, with unpolarized phase function $p_{\rm R}(\cos\theta)=3(1+\cos^2\theta)/(16\pi)$. In MC, absorption and scattering compete at rate $\mu_{\rm tot}+\mu_{\rm R}$; scattering changes direction but preserves wavelength. In the RTE, the corresponding loss $-\mu_{\rm R}I$ is balanced by the angular source $\mu_{\rm R}\int p_{\rm R}I\,d\Omega'$. Both incident and re-emitted photons are scattered. The Markov, 1D MC, and 2D MC calculations omit Rayleigh scattering.

The RTE uses a nonuniform Cartesian mesh, locally refined near the incident axis, absorption region, and collection windows. Across the tested sizes and excitation wavelengths, directional grid counts range from $10\times8\times22$ to $14\times20\times26$, or 1,760--7,280 cells. Cell widths span $0.0625$ mm--$1.25$ m in $x$, $0.0028$ mm--$1.25$ m in $y$, and $0.0625$ mm--$15$ m in $z$; shorter incident absorption lengths receive finer spacing near the source.

All spectra use 1-nm bins, and arrival-time densities use 0.25-ns bins over 0--150 ns. We use 350-nm excitation for the quantitative comparisons because Fig.~\ref{fig:all_size_comparison} shows the largest visible spectral separation between Markov and 3D MC at this excitation wavelength. Near this excitation wavelength, the initial F- and W-emission fractions are both appreciable and compete, so a modest increase in optical path length can increase secondary IFE and shift more F-origin emission toward W; this makes 350-nm excitation the worst case for the spectral comparison. Spectral total variation (TV) ranges from 0\% for identical spectra to 100\% for non-overlapping spectra, and the maximum time-CDF difference is evaluated over 0--150 ns on the same 0--100\% scale. The FF / RA results are summarized in Table~\ref{tab:cross_validation_metrics}; the 3D MC and RTE comparisons include Rayleigh scattering.

\begin{table}[htbp]
\centering\scriptsize
\caption{Spectral TV and maximum time-CDF difference for the worst-case spectral comparison at $350$-nm excitation. Entries give FF / RA in percent. The Markov--3D MC and RTE--3D MC comparisons include Rayleigh scattering.}
\label{tab:cross_validation_metrics}\label{tab:shape_scan}\label{tab:time_accuracy_all_sizes}
\setlength{\tabcolsep}{2.2pt}
\begin{tabular}{rcccccc}
\toprule
$L$ (m) & \multicolumn{3}{c}{Spectral TV (\%)} & \multicolumn{3}{c}{Maximum time CDF difference (\%)} \\
& Markov--1D MC & Markov--3D MC & RTE--3D MC & Markov--1D MC & Markov--3D MC & RTE--3D MC \\
\midrule
0.001 & 0.108 / 0.156 & 0.503 / 0.984 & 0.434 / 0.594 & 0.018 / 0.036 & 0.164 / 0.188 & 0.252 / 0.200 \\
0.01  & 0.111 / 0.158 & 3.054 / 3.396 & 0.400 / 1.122 & 0.015 / 0.017 & 2.041 / 0.479 & 0.189 / 0.142 \\
0.1   & 0.109 / 0.154 & 6.683 / 2.882 & 0.269 / 1.964 & 0.019 / 0.021 & 3.647 / 1.881 & 0.469 / 1.378 \\
1     & 0.111 / 0.165 & 6.702 / 1.888 & 0.395 / 1.586 & 0.011 / 0.016 & 3.657 / 7.228 & 0.683 / 6.918 \\
10    & 0.108 / 0.155 & 6.711 / 3.099 & 0.421 / 2.047 & 0.012 / 0.016 & 3.639 / 46.859 & 0.706 / 27.694 \\
\bottomrule
\end{tabular}
\end{table}

In Fig.~\ref{fig:all_size_comparison}, Markov and 1D Specific MC nearly coincide because the Markov model analytically solves their shared rectilinear transport. RTE and 3D MC also nearly overlap in the wavelength spectra and laboratory-scale time spectra; coincident curves can therefore obscure one another. Differences from isotropic 2D/3D results reflect the broader path-length measure: more long optical paths allow additional secondary IFE and shift F-origin emission toward W; Rayleigh scattering further redirects photons. For the tested millimetre- and centimetre-scale cuvettes, spectra and timing remain close to the isotropic calculations; at 10 m, timing is more sensitive than spectral shape to the broadened path-length distribution. These results support the prescribed-path Markov model for common laboratory cuvettes when interface reflection and refraction are suppressed and scattering is weak; larger volumes require geometry-resolved paths, especially for timing predictions.
 
\section{Computational Complexity and Performance}\label{sec:numerical_validation_complexity}

\subsection{RA Markov Complexity}\label{subsec:ra_complexity}

At each transform sample, wavelength integration forms the $2\times2$ transition matrix in Eq.~\eqref{eq:matrix_elements}; Eq.~\eqref{eq:matrix_equation} sums all generations, and Eq.~\eqref{eq:Psi_n_correct} gives the detector response. For RA, the collection response is evaluated at one prescribed distance, with no FF source-depth sweep. Let $N_\lambda$ be the wavelength-bin count, $K_p$ the upper spatial Talbot node count (the full contour has $2K_p-1$ nodes), $N_F$ the time-inversion FFT length, and $N_q=N_F/2+2$ the number of temporal transform samples used by the implementation. The leading theoretical FLOP counts for the RA spectrum and time spectrum are
\begin{equation}
\begin{aligned}
F_{\rm RA,spec}^{\rm lead}&=59K_pN_\lambda,
&\mathcal O(K_pN_\lambda),\\
F_{\rm RA,time}^{\rm lead}&=56K_pN_\lambda+160N_q(2K_p-1),
&\mathcal O\!\left[K_p(N_\lambda+N_F)+N_F\log N_F\right].
\end{aligned}
\label{eq:ra_flops}
\end{equation}
The steady spectrum needs one temporal value but projects onto all output wavelengths; the time response instead evaluates the transform across $N_q$ frequencies and adds the inverse FFT. The RA spectrum also uses a 32-point scalar source-depth quadrature, which contributes lower-order work. For $N_\lambda=401$, $K_p=24$, and $N_F=65{,}536$ ($N_q=32{,}770$), these theoretical leading counts are $0.568$ MFLOPs for the spectrum and $0.247$ GFLOPs for the time response. At these settings, the repeated time-frequency term $160N_q(2K_p-1)$ in Eq.~\eqref{eq:ra_flops} contributes about 99.8\% of the leading FLOP count, and its frequency samples are independent; parallelizing them therefore targets the dominant work before the inverse FFT. For the 350-nm, 1-mm RA time spectrum in Fig.~\ref{fig:all_size_comparison}, the mean runtime over 20 runs was 0.0416, 0.0238, 0.0148, and 0.0119 s with 1, 2, 4, and 8 CPU threads, respectively (speedups of $1.00$, $1.75$, $2.81$, and $3.51$). Because the serial calculation is already at the millisecond scale, the diminishing gains at higher thread counts are consistent with thread coordination and inter-core communication becoming increasingly significant relative to the computation itself.
\subsection{Computational Performance}\label{subsec:computational_performance}

For the implemented 3D RTE with Rayleigh scattering, the leading arithmetic cost is
\begin{equation}
\mathcal O\!\left[(J_\Sigma+N_t)N_rN_\Omega N_\lambda^{\rm R}
+P_\Sigma N_rN_\Omega\right],
\label{eq:rte_flops}
\end{equation}
Here $N_r$, $N_\Omega$, and $N_\lambda^{\rm R}$ are the spatial-cell, direction, and emission-wavelength counts; $N_t=1$ for a steady spectrum or 600 for the 0.25-ns time grid, and $J_\Sigma,P_\Sigma$ count fluorescence-field and incident-excitation updates. For 350-nm excitation at 1 cm, $(N_r,N_\Omega,N_\lambda^{\rm R})=(2640,32,300)$ and $(J_\Sigma,P_\Sigma)=(14,4)$ for the spectrum or $(1298,613)$ for the time calculation. Substitution in Eq.~\eqref{eq:rte_flops} with unit coefficient gives leading-scale estimates of $0.380$ and $48.2$ GFLOPs, respectively.

\begin{table}[htbp]
\centering\footnotesize
\caption{Calculation times for the 350-nm RA observables on an Intel Core i7-10700 CPU (2.90 GHz; Python 3.13.6, NumPy 2.4.4, SciPy 1.17.1). MC and Markov entries are mean $\pm$ sample standard deviation over 20 runs; RTE entries are the existing single-run times. All times are in seconds; Rayleigh-off values follow in parentheses.}
\label{tab:all_size_performance}
\setlength{\tabcolsep}{2pt}
\resizebox{\linewidth}{!}{%
\begin{tabular}{rccccccc}
\toprule
 & \multicolumn{3}{c}{MC time (s)} & \multicolumn{2}{c}{Spectrum time (s)} & \multicolumn{2}{c}{Arrival-time calculation (s)} \\
$L$ (m) & 1D Specific & 2D isotropic & 3D isotropic & Markov RA & 3D RTE & Markov RA & 3D RTE \\
\midrule
0.001 & 5.93$\pm$.20 & 6.64$\pm$.28 & 2.63$\pm$.13(2.48$\pm$.11) & 0.00109$\pm$0.00001 & 15.6(5.44) & 0.04096$\pm$0.00033 & 1909(881) \\
0.01 & 6.73$\pm$.19 & 7.66$\pm$.31 & 6.21$\pm$.18(5.87$\pm$.07) & 0.00109$\pm$0.00008 & 12.49(2.49) & 0.04098$\pm$0.00051 & 2664(1524) \\
0.1 & 8.14$\pm$.22 & 8.47$\pm$.41 & 7.11$\pm$.23(6.68$\pm$.10) & 0.00108$\pm$0.00001 & 50.29(13.08) & 0.04099$\pm$0.00052 & 4383(3406) \\
1 & 8.97$\pm$.29 & 8.81$\pm$.39 & 7.37$\pm$.14(6.98$\pm$.12) & 0.00108$\pm$0.00001 & 51.33(14.41) & 0.04065$\pm$0.00036 & 4387(3404) \\
10 & 9.65$\pm$.25 & 9.06$\pm$.36 & 8.06$\pm$.14(7.19$\pm$.11) & 0.00108$\pm$0.00008 & 51.02(14.61) & 0.04134$\pm$0.00040 & 4385(3400) \\
\bottomrule
\end{tabular}
}
\end{table}

Table~\ref{tab:all_size_performance} summarizes the measured runtimes. At $L=1$ cm, the single-thread RA-only Markov calculation takes 0.00109 s for the spectrum and 0.0410 s for the arrival-time response, compared with 6.21 s for 3D MC; the measured Markov-to-CPU-MC speedups are $5.7\times10^3$ and $1.5\times10^2$, respectively. As a throughput reference, the NEXT Opticks study reports GPU speedups of 58.47--181.39$\times$ over its tested CPU configurations \cite{next_collaboration_performance_2025}. The Markov spectral speedup exceeds this reported range, while its time-spectrum speedup is near the upper end. At the stated MC history budget, low-probability late arrivals remain poorly sampled, whereas the Markov calculation evaluates these contributions directly. The 3D RTE takes 12.49 s for the spectrum and 2664 s for the arrival-time response. Together with the agreement in Sec.~\ref{subsubsec:mc_validation}, these results support the Markov model as a fast forward calculation for repeated optical-parameter tuning in the tested laboratory-cuvette range.

\section{Conclusion}\label{sec:conclusion}

This paper has established a joint spatiotemporal transport model that maps convoluted multi-generation photon cascades onto an algebraic Markovian transition matrix recurrence equation via a Laplace transform, reducing the cost of direct nested wavelength integration from $\mathcal{O}(N_{\lambda}^n)$ to $\mathcal{O}(N_{\lambda} + n)$ per transform evaluation at fixed component number. By taking the spatial marginalization limit ($p \to 0^+$), the joint formulation reduces to the pure time domain, characterizing intrinsic emission-event time profiles as a continuum superposition of wavelength-dependent Gamma wave packets for exponential microscopic delays. The joint finite-path response defines a regularized inversion strategy for laboratory signals. Extending this mathematical system to the steady-state temporal limit ($s \to 0$) yields continuous wavelength spectrum predictions and establishes explicit, closed-form correction formulas across absorbance, right-angle, and front-face spectrometers. The sub-Markovian analysis supplies convergence conditions and finite-order truncation bounds. Under the stated non-amplifying conversion condition, the energy balance proves mean-energy conservation for the photon--medium system; if $\rho(\mathbf M_0)<1$, the residual photon energy vanishes and the cascade is dissipative.

Although the transport equations accommodate specific propagation distances, the analytical model does not automatically track reflections and refractions at optical boundaries or isotropic random walks. Comparisons with 1D MC verify the analytical reduction, while isotropic MC and RTE, with and without Rayleigh scattering, support good spectral and timing agreement for common laboratory cuvettes under the stated collection assumptions. At large dimensions, timing is more sensitive to the broader distribution of optical path lengths. With millisecond spectrum and tens-of-milliseconds time-spectrum calculations, both faster than RTE, the main utility of this analytical model is established as a fast, physics-preserving forward calculation tool to guide and accelerate parameter optimization across multiple experimental setups.

Potential applications of this analytical model include:
\par
\begingroup
\leftskip=2.5em\relax
\parindent=0pt\relax
\noindent\llap{\textbullet\hspace{0.5em}}\textit{Optical Parameter Correction}: Screening vast parameter spaces to simplify parameter extraction. For $N$ variables evaluated across 400 discrete wavelengths, the resulting $400 \times N$ parameter space can require substantial computational time via repeated brute-force MC. The algebraic solution isolates candidate parameter regions before performing final verifications with MC.\par
\noindent\llap{\textbullet\hspace{0.5em}}\textit{Phenomenological Parameter Extraction}: Isolating intrinsic material variables from raw measurements by algebraically removing the spectral distortions and temporal broadening caused by the secondary IFE. This method enables the phenomenological quantification of traditionally unmeasurable parameters, such as non-radiative transfer probabilities and times under alternative $dE/dx$.\par
\noindent\llap{\textbullet\hspace{0.5em}}\textit{Fast Detector Design Reference}: Accelerating early-stage detector designs for alternative fluorescent mixtures and geometric dimensions. This approach bypasses the extensive time required to construct detailed geometry simulations for initial design evaluations.\par
\noindent\llap{\textbullet\hspace{0.5em}}\textit{Vertex Reconstruction Input}: Serving as a physics-constrained forward input for event vertex reconstruction algorithms. The physical model provides a theoretical reference for empirical neural network emulators.\par
\endgroup

\begin{appendices}
\renewcommand{\thesection}{\arabic{section}}

\section{Competing Kinetics and Time-Dependent Branching}\label{sec:appendix_competing_kinetics}

\textbf{This appendix extends the matrix formulation to nonconstant competing rates and states the condition under which the extension applies.} After excitation enters component $j$, it can undergo non-radiative transfer to $j+1$, radiative emission from $j$, or quenching in $j$, at rates $k_{\rm nr}^{j\to j+1}(t)$, $k_r^j(t)$, and $k_q^j(t)$. Their sum is $k_\Sigma^j(t)$. The survival probability $U^j(t)=\exp[-\int_0^t k_\Sigma^j(u)\,du]$ is the probability that none has yet occurred. Its negative derivative, $-dU^j/dt=k_\Sigma^j(t)U^j(t)$, is the waiting-time density for whichever of these three events—transfer, emission, or quenching—occurs first.

At elapsed time $t$, $P_{\rm nr}^{j\to j+1}(t)=k_{\rm nr}^{j\to j+1}(t)/k_\Sigma^j(t)$ is the conditional fraction of transfer events among transfer, emission, and quenching; conditional on no transfer, $Q^j(\lambda,t)=k_r^j(t)/[k_r^j(t)+k_q^j(t)]$ is the fraction of emissions among emission and quenching. The total waiting-time density $k_\Sigma^j(t)U^j(t)$ sets when one of these outcomes occurs. Multiplying it by the corresponding conditional fractions gives the event-time densities for transfer, emission, and quenching, respectively: $P_{\rm nr}^{j\to j+1}(t)k_\Sigma^j(t)U^j(t)=k_{\rm nr}^{j\to j+1}(t)U^j(t)$, $[1-P_{\rm nr}^{j\to j+1}(t)]Q^j(\lambda,t)k_\Sigma^j(t)U^j(t)=k_r^j(t)U^j(t)$, and $[1-P_{\rm nr}^{j\to j+1}(t)][1-Q^j(\lambda,t)]k_\Sigma^j(t)U^j(t)=k_q^j(t)U^j(t)$. The transfer and emission densities are exactly the two factors transformed in $\mathcal A_k$ below. Quenching has no photon-output term, but $k_q^j$ enters the exponent of $U^j$ and therefore reduces both $k_{\rm nr}^{j\to j+1}U^j$ and $k_r^jU^j$. Assumption~4 makes successive waiting times independent when transfer starts the next component's clock. Replacing the branch-weighted PDFs in Eq.~\eqref{eq:phi_unified_time} by these time-dependent transfer and emission densities gives the corresponding coefficient $\mathcal A_k$, in place of Eq.~\eqref{eq:A_components_generalized}:
\begin{equation}
\begin{aligned}
\mathcal A_k(\lambda,s)
&=\left[\sum_{i=1}^k\mu_i(\lambda)\prod_{j=i}^{k-1}
\mathcal L\{P_{\rm nr}^{j\to j+1}(t)k_\Sigma^j(t)U^j(t)\}(s)\right]\\
&\quad\times\mathcal L\{[1-P_{\rm nr}^{k\to k+1}(t)]Q^k(\lambda,t)
k_\Sigma^k(t)U^k(t)\}(s).
\end{aligned}
\label{eq:A_time_dependent_branches}
\end{equation}
The sum selects the absorbing component and the product its successive transfers up to emission from $k$. Since output wavelength enters only through $S_{\rm em}^k$, the matrix equations retain their form.

\section{Limits of Relative Deconvolution}\label{sec:appendix_deconvolution_limits}

\textbf{Relative measurements cancel a shared instrument response; this appendix tests whether they also cancel IFE.} Following Eq.~\eqref{eq:Psi_n_correct}, define the final-flight detector factor $D_k(s,p)=\int_0^\infty S_{\rm em}^k(\lambda)QE(\lambda)/[p+\mu_{\rm tot}(\lambda)+s/v]\,d\lambda$. Section~\ref{subsec:Sub-Markov} permits summing the generation series, $\sum_{n\ge0}\boldsymbol{\mathcal M}^n=(\mathbf I-\boldsymbol{\mathcal M})^{-1}$. At fixed optical path length $d$, the response to a component source $\vec C_{\rm start}(s)$ is $\widetilde\Psi_{\rm det}(s;d)=\vec{\mathcal H}(s;d)^T\vec C_{\rm start}(s)$, where
\begin{equation}
\vec{\mathcal H}(s;d)^T=\mathcal L^{-1}_{p\to d}\!\left\{\vec D_{\rm det}(s,p)^T[\mathbf I-\boldsymbol{\mathcal M}(s,p)]^{-1}\right\}.
\label{eq:appendix_B3}
\end{equation}
Each entry of $\vec{\mathcal H}$ is the detected response per unit initial emission from one component, including its spectrum, subsequent IFE, and final flight; these entries generally differ across components. For two sources sharing the collection geometry and instrument kernel, $\widetilde S^{(r)}=\widetilde{\mathcal K}_{\rm inst}\widetilde\Psi_{\rm det}^{(r)}$ gives, wherever the denominator is nonzero,
\begin{equation}
\frac{\widetilde S^{(1)}(s)}{\widetilde S^{(2)}(s)}
=\frac{\widetilde\Psi_{\rm det}^{(1)}(s;d)}{\widetilde\Psi_{\rm det}^{(2)}(s;d)}
=\frac{\vec{\mathcal H}(s;d)^T\vec C_{\rm start}^{(1)}(s)}
{\vec{\mathcal H}(s;d)^T\vec C_{\rm start}^{(2)}(s)}.
\label{eq:appendix_B5}
\end{equation}
Cancellation for arbitrary component responses requires collinear sources: if $\vec C_{\rm start}^{(1)}=\zeta\vec C_{\rm start}^{(2)}$, the ratio in Eq.~\eqref{eq:appendix_B5} reduces to $\zeta$. The source-specific fractions and delays in Eq.~\eqref{eq:initial_state_dedx_correct} generally give non-collinear populations, so IFE remains in the ratio; a small denominator also amplifies numerical errors.

\end{appendices}

\begin{backmatter}
\bmsection{Funding}
National Natural Science Foundation of China (12375196, 12275293).

\bmsection{Disclosures}
The authors declare no conflicts of interest.

\bmsection{Data Availability}
Data underlying the results presented in this paper are not publicly available at this time but may be obtained from the authors upon reasonable request.

\bmsection{Declaration of Generative AI Technologies} 
During the preparation of this work, the author utilized the Google Gemini 3.1 and OpenAI Codex with GPT-6 to assist in the paper research. Under the conceptual framework and theoretical design established solely by the author, this tool was employed to verify the self-consistency of specific algebraic derivations, implement the underlying numerical models into Python and MATLAB code, and refine the technical text into standard academic English. Following the use of this service, the author reviewed and verified all mathematical derivations, algorithmic implementations, and textual contents, and takes full responsibility for the content and conclusions of the published article.
\end{backmatter}

\bibliography{Manuscript_References}      

\end{document}